\documentclass[aps, prl, twocolumn]{revtex4-1}
\usepackage{hyperref}
\usepackage{bm}
\usepackage{amsmath}
\usepackage{amssymb}
\usepackage{graphicx}
\usepackage{tabularx}
\usepackage{tabu}
\usepackage{longtable}
\usepackage{multirow}
\usepackage{color}
\usepackage[table]{xcolor}
\usepackage{pstool}
\usepackage[absolute,overlay]{textpos}
\usepackage{tcolorbox}
\hypersetup{
colorlinks=true,
citecolor=blue,
urlcolor=blue
}

\newcommand{\Tr}{\text{Tr}}
\newcommand{\sinc}{\text{sinc}}

\definecolor{color1}{RGB}{230,230,250}
\definecolor{color2}{RGB}{255,255,224}
\definecolor{green2}{RGB}{0,150,0}

\newcommand{\newc}{\newcommand}
\newc{\beq}{\begin{equation}}
\newc{\eeq}{\end{equation}}
\newc{\kt}{\rangle}
\newc{\br}{\langle}
\newc{\beqa}{\begin{eqnarray}}
\newc{\eeqa}{\end{eqnarray}}
\newc{\longra}{\longrightarrow}
\newc{\Ob}{{\mathcal O}}

\begin{document}

\title{Scrambling in strongly chaotic weakly coupled bipartite systems: Universality beyond the Ehrenfest time-scale}
%\title{Out-of-time-order correlators in bipartite chaotic systems: universal and non-universal features}
\author{Ravi Prakash} \email{raviprakash.sps@gmail.com}
\author{Arul Lakshminarayan}\email{arul@iitm.ac.in}
\affiliation{Indian Institute of Technology Madras, Chennai -- 600036, India}
\begin{abstract}
Out-of-time-order correlators (OTOC), vigorously being explored as a measure of quantum chaos and information scrambling,
is studied here in the natural and simplest multi-particle context of bipartite systems. We show that two strongly chaotic
and weakly interacting subsystems display two distinct phases in the growth of OTOC. 
The first is dominated by intra-subsystem scrambling, when an exponential growth with a positive Lyapunov exponent is observed 
till the Ehrenfest time. This phase is essentially independent of the interaction, while the second phase is an interaction 
dominated exponential approach to saturation that is universal and described by a random matrix model.
This simple random matrix model of weakly interacting strongly chaotic bipartite systems, previously employed for studying entanglement and spectral transitions, is approximately analytically solvable for its OTOC. The example of two coupled kicked rotors is used to demonstrate the two phases, and the extent to which the random matrix model is applicable. That the two phases correspond to delocalization in the subsystems 
followed by inter-subsystem mixing is seen via the participation ratio in phase-space. We also point out that the second, universal, phase alone exists when
the observables are in a sense already scrambled.
Thus, while the post-Ehrenfest time OTOC growth is in general not well-understood, the case of strongly chaotic and weakly coupled systems presents an, perhaps important, exception. 
\end{abstract}
\maketitle
%\section{Introduction}
% write about scrambling time.
The quantum mechanics of classically chaotic, or in general non-integrable, systems known generically as quantum chaos presents a complex of interesting features. The universal spectral fluctuations of quantum chaotic systems are widely studied using random matrix theory (RMT) \cite{Mehta, Akemann-RMT, Porter, Brody-1981}, while semiclassical spectra are described via periodic orbit theories \cite{Gutzwiller-1970, Gutzwiller-1990}. A large part of the previous works studied either time-evolving states or eigenspectra \cite{Cerruti-2003, Gorin-2004}, while a recent trend concerns operator evolution and is therefore tied in more directly to the evolution of classical observables. Operator spreading or scrambling and out-of-time-ordered correlators (OTOC) are two quantities on which much attention has been bestowed from diverse areas \cite{Shen-2017, Slagle-2017, Cotler-2017, Campisi-2017, Hashimoto-2017, Fan-2017, Keyserlingk-2018, Nahum-2018, Rozenbaum-2017, Jalabert-2018, Arul-Baker-2018, Rakovszky-2018, Saraceno-2018, Chen-2018, Moudgalya-2018, Haehl-2017}.  

OTOC, introduced in the context of superconductivity \cite{Larkin-1969}, is now being widely studied in many contexts, such as quantum gravity field theories and many-body physics, including many-body localization, models such as random quantum circuits as well as  quantum walks and weak measurements  \cite{Fan-2017,  Chen-2017, Slagle-2017, Keyserlingk-2018, Omanakuttan-2019, Halpren-2019}. The OTOCs have been recently related to the other measures of quantum chaos such as spectral statistics, participation ratio and Loschmidt echo \cite{Borgonovi-2019, Yan-2019, Rozenbaum-2019, Rozenbaum-2018}. Thus the OTOC provides a window far beyond conventional settings of quantum chaos besides providing an opportunity for new quantum measures of complexity. In particular, for chaotic systems the OTOC grow exponentially till the Ehrenfest time \cite{Berry-1979, Berry-1979_2} providing a quantum Lyapunov exponent \cite{Rozenbaum-2017, Maldacena-2016}. 

The correspondence is most transparent in the increase of non-commutativity of two (say Hermitian) operators, one evolving  with time. 
Consider
\begin{equation}
\label{eq-otoc}
C(t) = -\frac{1}{2} \left< [A(t),B(0)]^2 \right>,
\end{equation}
where $\left< \cdot \right>$ represents the thermal average over an ensemble at temperature $T$. A standard semiclassical argument makes
it plausible that this can increase exponentially with time. If $A$ and $B$ are the position and momentum operator the commutator $[x(t), p(0)]^2$ is semiclassically the Poisson bracket, $\hbar^2 \{x(t),p(0)\}^2 = \hbar^2(\partial x(t)/\partial x(0))^2$, which grows exponentially for chaotic systems, as $(\partial x(t)/\partial x(0))^2 \approx \exp(2 \lambda t)$ due to the sensitive dependence on initial conditions.
 It is argued that the rate has an upper bound, $\lambda \leq 2\pi k_B T/\hbar$ \cite{Maldacena-2016}. The Sachdev-Ye-Kitaev (SYK) model, a disordered model of Fermions with all-to-all interactions, is one of the maximally chaotic system which saturates the bound \cite{SYK-1993, Kitaev-2015}. Similar bound was found in earlier studies of scrambling of quantum information around a black hole horizon \cite{Shenker-2014, Preskill-2007}.

The exponential growth of $C(t)$, the Lyapunov phase, occurs in a time window $t_d<t<t_{EF}$ where $t_d$ is a diffusion time scale that is comparatively small and does not scale with the system size, while $t_{EF}$ is the Ehrenfest time and could be the time of breakdown of classical-quantum correspondence if a classical limit exists. There have now 
been several studies of the OTOC on models of quantum chaos, such as the quantum standard map, the quantum bakers map, the cat map, and the
kicked top, and on all-to-all connected spin models \cite{Rozenbaum-2017, Cotler-2018, Chen-2018, Saraceno-2018, Hamazaki-2018}. All these display the expected exponential growth till the Ehrenfest time,  which scales as $\sim \log N$, where $N$ is the Hilbert space dimension. 

 Beyond the Ehrenfest time, the $\hbar$ corrections start dominating \cite{Cotler-2018}, and there exists no classical correspondence for the OTOC, even if the system has a well-defined semiclassical limit, marking a less understood phase that is important to study in various settings. This Letter is concerned with the simplest multipartite system, a generic bipartite one given by $H = H_1 \otimes \mathbb I_2 + \mathbb I_1 \otimes H_2 + b\,V_{12}$
where $H_j$ are strongly chaotic subsystem Hamiltonians and $V_{12}$ is an interaction which is kept small by requiring the dimensionless $b \ll 1$.

Consider $A(0)$ and $B(0)$ to be localized to either subsystem. For the most part we will be concerned with the interesting case when they are localized on different subsystems:  $A(0) = \Ob_1\otimes \mathbb{I}_2,\; B(0) = \mathbb{I}_1 \otimes \Ob_2$.
The Heisenberg evolution of operator $A(0)$ renders it entangled and therefore $A(t)$ fails to commute with $B(0)$ for $t>0$.
Thus this is the simplest multipartite setting in which entanglement is responsible for the OTOC growth and information scrambling. We find that if 
the operators $A(0)$, $B(0)$ have smooth classical limits, there are two
distinct epochs, the first being one of exponential growth, the Lyapunov phase, lasting for the Ehrenfest time of the subsystems. This epoch is  dominated by {\it intra-subsystem} scrambling and the Lyapunov exponent is largely independent of the interaction strength $b$. 

\begin{figure}[!h]
\centering
\includegraphics[scale=.5]{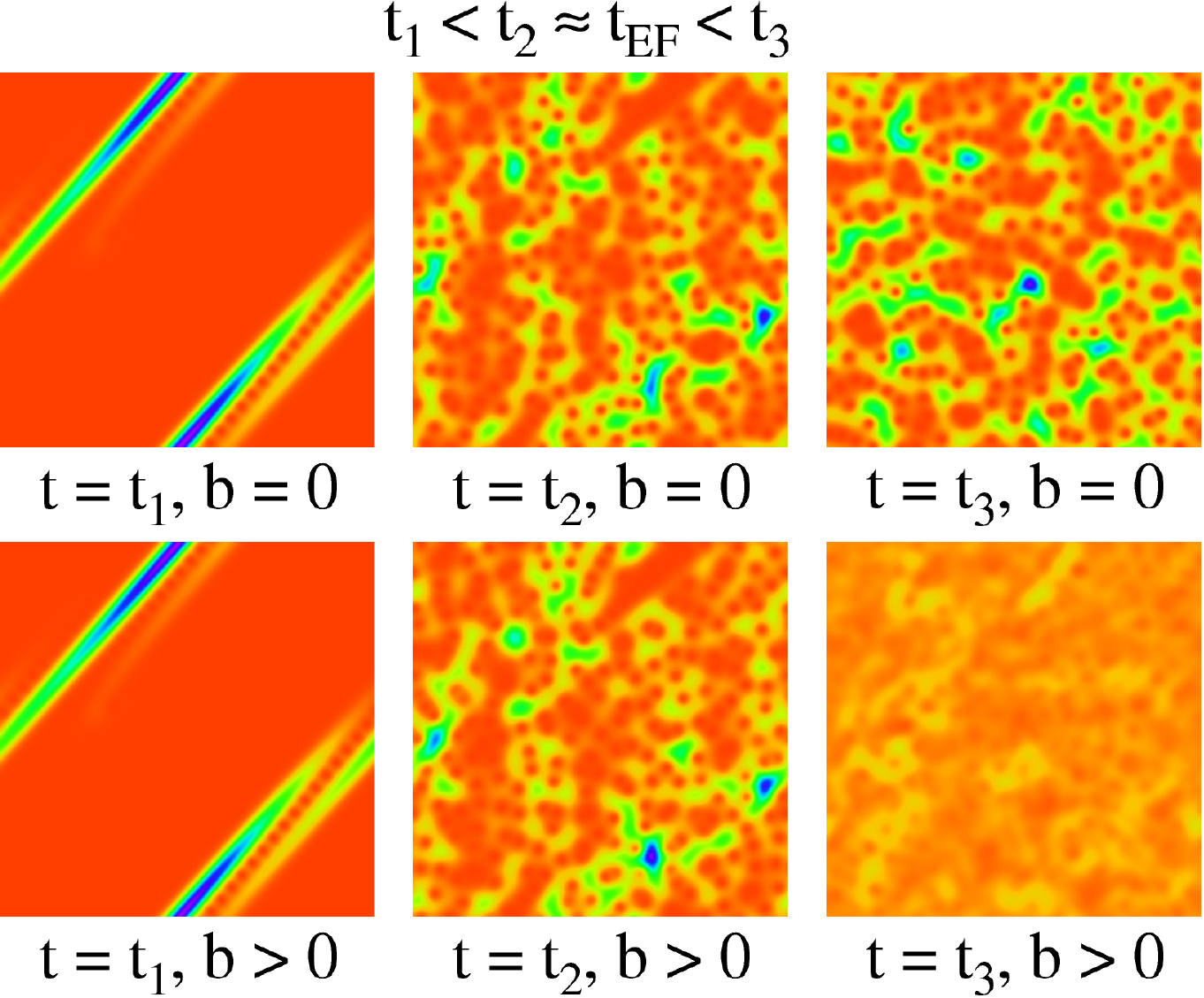}
\caption{(color online) Illustrating intra- and inter- subsystem scrambling. Shown are phase-space representations of initially localized time evolving states in one subsystem when both the subsystems are strongly chaotic. The top row 
is the case of no interaction ($b=0$), while the bottom row has $0<b \ll 1$. 
Till the Ehrenfest time $t_{EF}$, the state in both cases are essentially the same and get scrambled within the subsystem, while after $t_{EF}$, the subsystem state decoheres significantly due to inter-subsystem scrambling.
 \label{fig:schemat}}
\end{figure}

The second epoch is one of exponential relaxation with  a rate that is strongly interaction dependent. This marks an era of {\it inter-subsystem} scrambling and is universal in some sense, the rate being well predicted by a random matrix model that we develop here. Fig.~(\ref{fig:schemat})
illustrates this scenario in the phase space evolution of localized densities. As a quantitative measure we also show that the participation ratio in phase space, a measure of its delocalization exhibits a clear difference between the intra- and inter- subsystem scrambling phases.

If the observables used are not smooth operators, or are
in some sense {\it pre-scrambled} the Lyapunov phase can be entirely absent and the OTOC relaxes exponentially from the start with the universal rate. It is interesting that even in the absence of the Lyapunov phase, it is possible to distinguish a chaotic system from non-chaotic ones, as we provide some numerical evidence that in the latter case, a putative saturation is approached algebraically ($\sim 1/t$) rather than exponentially. 

To begin with, we study a concrete dynamical system, two coupled kicked rotors,
illustrating the main features, and later use RMT to derive the exponential relaxation rate.
Rather than using thermal averages, we consider the infinite temperature limit, the OTOC for two operators $A$ and $B$ given in Eq.~(\ref{eq-otoc}) is then $C(t)  =  C_2(t) - C_4(t),$ where
\begin{equation}
\label{eq-otoc-2}
C_2(t) = \text{Tr}\left[A(t)^2 B(0)^2\right], \;
C_4(t) = \text{Tr}\left[A(t)B(0)A(t)B(0)\right] 
\end{equation}
and $C_4(t)$ is an 4-point out-of-time ordered correlator.

\noindent{\it OTOC for coupled quantum kicked rotors}:
 A rich, yet simple, class of models results when the subsystems Hamiltonians 
$H_j$ are 1-degree of freedom periodically forced systems. A well-studied paradigmatic model is that of two coupled kicked rotors \cite{Wang-1990, Wood-1990,Arul-2001,Richter-2014} , for which
$H_j = \frac{1}{2}p_j^2 +  \frac{1}{4\pi^2} K_j \cos(2\pi q_j)\delta_t$ , and $ b V_{12} =  \frac{b}{4\pi^2} \cos(2\pi(q_1 + q_2))\delta_t$,
where $\delta_t = \sum_{n = -\infty}^{\infty} \delta(t-n)$. The parameter $b$ is the interaction while the individual rotor parameters $K_j$ determine local chaos. The single rotor is integrable only for vanishing kick strengths $K_j = 0$, and there is a mixed phase space, with a finite measure of chaotic and stable regions as $K$ increases, with widespread chaos for $K \gg 5$.

As is well-known, the quantum dynamics of the kicked rotors with  torus boundary conditions occurs in a finite dimensional Hilbert space of dimension say $N$, so that both position and momentum have discrete values. The Hilbert space of two coupled rotors is the tensor product space of dimension $N^2$ on which the Floquet operator is of the form
\begin{equation}
\label{eq-floquet}
U = (U_{K_1} \otimes U_{K_2}) U_b,
\end{equation}
where $U_{K_j}=\mathcal T e^{-i \int_0^1 H_j dt /\hbar}$ are Floquet operators of individual rotors and $U_b=e^{-ib V_{12}/\hbar}$ is the interaction, explicit expressions are in \cite{supplemental}.
Since position and momentum are both discrete, it is convenient to have local observables to be constructed from their translation operators $T_q$ and $T_p$, $T_q \left | q_n \right > = \left | q_{n+1} \right >$ and $T_p \left | p_n \right > = \left | p_{n+1} \right >$. In particular
the observables we study are locally simply $\Ob = \frac{1}{2} (T_p + T_p^\dagger)$ with $\left<n\left |T_p\right | n^\prime \right > = \exp\left[2\pi i (n+\alpha)/N\right] \delta_{n n^\prime}$ in position basis, the classical limit being $\cos(2 \pi q_j)$.
We consider cases when the two rotor parameters are large and hence the subsystems are strongly chaotic.

\begin{figure}[!h]
\centering
\includegraphics[width=\linewidth]{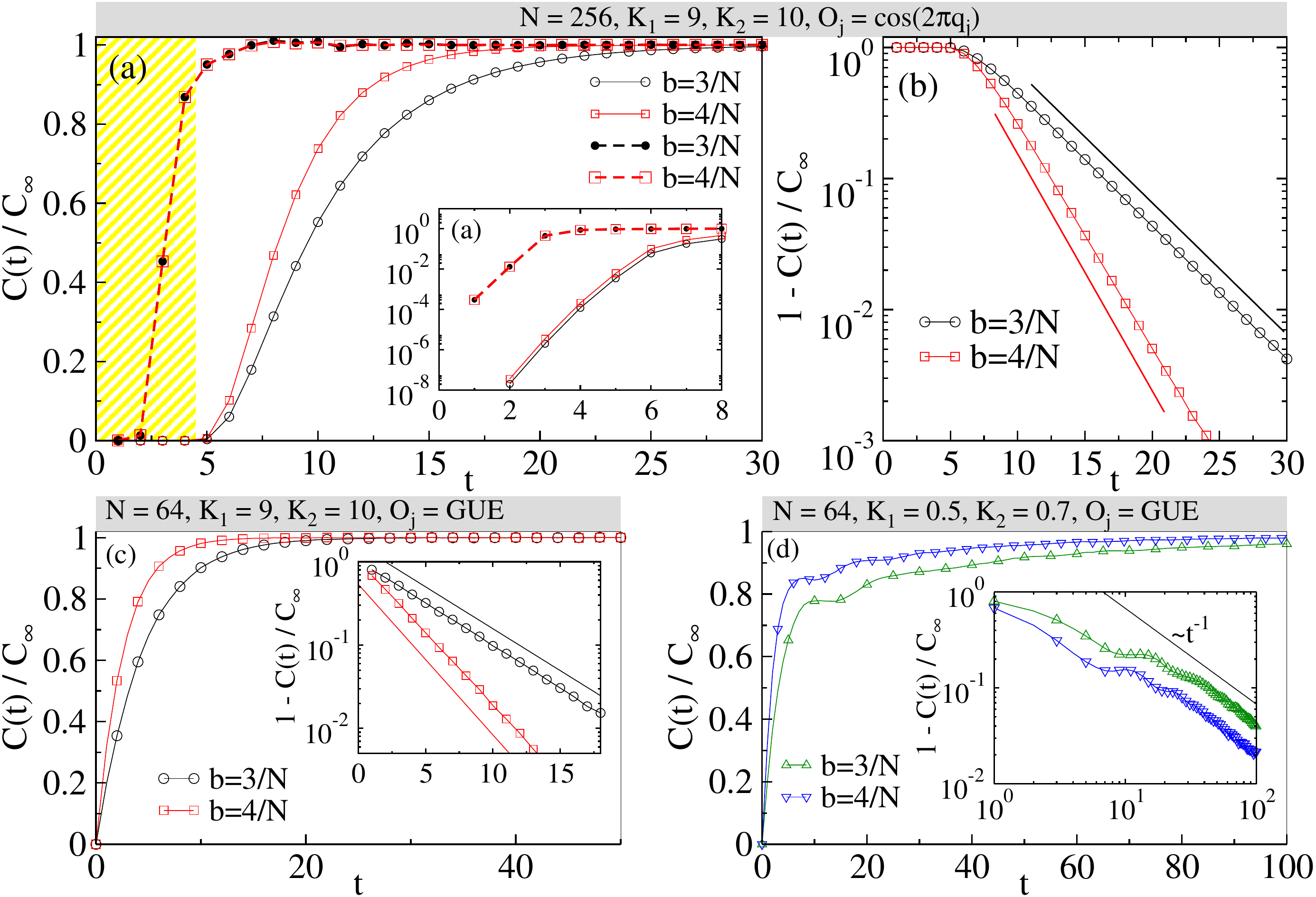}
\caption{(color online) OTOC growth with time is shown for various cases. The kick parameters are $K_1 = 9, K_2 = 10$ for parts (a), (b) and (c). (a) Normalized OTOC $C(t)/C_{\infty}$ for two cases, operators in different subspace (solid lines) and in same subspace (dashed lines). Inset shows the exponential growth till Ehrenfest time in the initial growth (log-linear scale), this is the Lyapunov phase with intra-subsystem scrambling. (b) $(1-C(t)/C_\infty)$ (log-linear scale) for operators in different subspace. The solid lines represents the rate given in Eq.~(\ref{eq:standmaprate}), and illustrates the RMT phase with inter-subsystem scrambling. (c) Illustrates the use of a random observable that has no Lyapunov, but only an RMT phase. (d) Random observables in a weakly chaotic system, the inset supports an  $1/t$ approach to saturation.\label{fig-otoc-fft}}
\end{figure}

The OTOC $C(t)$ from Eq.~(\ref{eq-otoc-2}), with $A(0)=\Ob_1 \otimes \mathbb I_2$ and $B(0)=\mathbb I_1 \otimes \Ob_2$ and $U$ the coupled standard map from Eq.~(\ref{eq-floquet}), is shown in Fig.~(\ref{fig-otoc-fft}). 

The uncoupled subsystems are highly chaotic and nearly identical.
The interactions, being only  of the order $1/N$ implies that measures of chaos
such as the Lyapunov exponent $\lambda$ is that of the uncoupled systems $\approx \ln(K_2/2)$.
There are clearly two phases, first is an exponential growth corresponding to the Lyapunov phase till the Ehrenfest time
corresponding roughly to that of the uncoupled subsystems $t_{EF} \approx \ln N/\lambda$. This phase is dominated by the sub-system chaos and the rate of exponential
growth is to a good approximation independent of the interaction. This would be the regime of intra-subsystem scrambling with the coupling determining
only the prefactor of the OTOC which is $C(t)\approx \gamma_1(b) e^{2 \lambda_{\text{cl}} t}$. Here $\lambda_{\text{cl}}$ comes from a classical Poisson bracket evaluation of $\{ \cos(2 \pi q_1(t)), \cos(2 \pi q_2(0)\}^2 \sim e^{2 \lambda_{\text{cl}} t}$, and is systematically larger than $\lambda$ \cite{supplemental}. 
The interactions although classically small, are large enough that the stationary state properties of the coupled system are that of random matrices. There is a  dimensionless transition parameter $\Lambda \sim N^4 b^2$ \cite{Arul-PRL-2016} which is such that if $\Lambda \gg 1$, the nearest neighbor spacing statistics is Wigner and 
the eigenfunctions are highly entangled. Thus we are already in the strong coupling regime as far as the stationary state properties are concerned. 

That this first phase is dominated by intra-subsystem scrambling is further evidenced by the OTOC between observables in the same subsystem. Thus if $A(0)=B(0)=\mathcal{O}_1 \otimes \mathbb{I}_2$,
the OTOC grows only the first phase and already saturates without a second phase as shown in Fig.~(\ref{fig-otoc-fft}a). Different interaction strengths do not affect the growth significantly, and we turn to the more interesting case of the post-Ehrenfest growth of the OTOC when the observables are local on different subsystems.

In this case. the second phase is a slower exponential relaxation to the saturation value during which the relaxation rate is strongly interaction 
dependent and practically independent of the subsystem parameters $K_j$. This inter-subsystem scrambling is responsible for the entanglement and eventually leads to random states on the product Hilbert space. This phase is universal in being only dependent on the fact that the subsystems are strongly chaotic and therefore is amenable to a random matrix treatment. Thus the numerical results support an approximate OTOC:
\beq
\label{eq:OTOCform}
C(t)=\left\lbrace \begin{array}{lr} \gamma_1(b) e^{2 \lambda_{cl} t}, & 0 <t \leq t_{EF} \\ C_{\infty} - \gamma_2(b) \, e^{-\mu(b) (t-t_{EF})} & t > t_{EF}, \end{array} \right.
%C(t)=\gamma(b) e^{2 \lambda t} \Theta(t_{EF}-t) + C(\infty) (1-e^{-\mu (t-t_{EF})})\Theta(t-t_{EF})
\eeq
here $C_{\infty}$ and $\gamma_2(b)$ are independent of time.  We now turn to deriving this relaxation based on random matrix theory, in particular we show that for the standard map discussed above 
\beq
\label{eq:standmaprate}
\mu(b)= \ln \left|J_0\left(\frac{N b}{2\pi} \right)\right |^{-4} \approx \frac{N^2 b^2}{4 \pi^2},
\eeq
where $J_0(x)$ is a Bessel function and this is valid for $Nb \ll 2 \pi$. The exponential relaxation in the second phase is shown in 
Fig.~(\ref{fig-otoc-fft}b) along with lines of this slope showing that this is a good approximation even for relatively  large values of the coupling, till $b \approx 3/N$ \cite{supplemental}. The saturation value $C_{\infty}=\Tr(\Ob_1^2) \Tr(\Ob_2^2)$ is obtained presently from an RMT analysis.

\noindent{\it Pre-scrambled operators}: While the second phase is universal and independent of the observables, the first phase can be completely absent if the observables do not have a smooth classical equivalent, say through the Weyl-Wigner symbol. An extreme case of this, could be termed as a pre-scrambled operator, which has fluctuations at the scale of $\hbar$ and is already ergodic in some sense. We take realizations of Gaussian random matrices as the local observables, $ \mathcal O=(M+M^{\dagger})/2$, where $M$ is a complex matrix, whose entry's real and imaginary parts are i.i.d. Gaussian random numbers with $0$ mean and unit variance, in other words from the GUE ensemble \cite{Mehta, Akemann-RMT}.

Fig.~(\ref{fig-otoc-fft}c) shows that the log-time growth is absent and that the relaxation is well described by the second part of Eq.~(\ref{eq:OTOCform}) with the rate given by Eq.~(\ref{eq:standmaprate}). The role of subsystem chaos in the second phase is to lead to an exponential relaxation. If the subsystems were not chaotic, say $K_1=0.5$ and $K_2=0.7$, the second phase with GUE observables shows a clear algebraic approach to saturation and numerical results support a $1/t$ approach as shown in Fig.~(\ref{fig-otoc-fft}d). It maybe noted that integrable spin chains have been observed to have such a behavior \cite{Lin-2018, Bao-2019}, and we postpone the study of the rich and complex scenario of mixed phase-spaces, turning to an analytical treatment of the strongly chaotic cases. 

\noindent {\it OTOC in a bipartite RMT model}:
In the case of strong subsystem chaos, the form of the Floquet operator in Eq.~(\ref{eq-floquet}) motivates replacing the local unitary maps $U_{K_j}$ with random unitary matrices, and for analytical tractability it is expedient and useful to take these as independent at different time steps. Thus, we take for the powers $\mathcal U^t$ the ensemble 
\begin{equation}
\label{eq-Ut}
\mathcal U^{(t)} =  \prod_{j=1}^t(\mathcal F_{1j} \otimes \mathcal F_{2j}) \mathcal U_{j\epsilon},
\end{equation}
where the $\mathcal F_{1j}$ and  $\mathcal F_{2j}$ are 
independent realizations from the circular unitary ensemble, CUE
that samples matrices uniformly from the group $U(N)$. The interaction is taken as a random diagonal matrix  
$\br m_1, n_1 \left|\mathcal{U}_{j \epsilon}\right|m_2, n_2\kt =\exp(2\pi i \epsilon \xi_{m_1n_1}) \delta_{m_1 m_2}  \delta_{n_1 n_2}$, where $\xi_k$ are uniform random in $[-1/2,1/2]$ and independent for each time $j$.
It has been shown that as $\epsilon$ increases from $0$, there is a transition in nearest neighbor level spacing statistics from an uncorrelated Poisson to the Wigner distribution \cite{Arul-PRL-2016} and this is accompanied by a universal transition in eigenstate entanglement from $0$ to the nearly maximal random state average \cite{Arul-2018}. We now explore this in the time domain mainly via the OTOC, but also via participation ratio in phase space.

%\section{Out of time order correlator} \label{sec-otoc}

With a view towards deriving a recursive scheme, write the four point function in Eq.~(\ref{eq-otoc-2}) as,
\begin{eqnarray}
\label{eq-c4}
\nonumber C_4(t) & =\Tr \left[ \mathcal U_{t \epsilon}^\dagger (\mathcal F_{1t}^\dagger \otimes \mathcal F_{2t}^\dagger) A(t-1) (\mathcal F_{1t} \otimes \mathcal F_{2t}) \mathcal U_{t \epsilon} B(0) \right.\\
&\left. \times \mathcal U_{t \epsilon} ^\dagger (\mathcal F_{1t}^\dagger \otimes \mathcal F_{2t}^\dagger) A(t-1) (\mathcal F_{1t}\otimes \mathcal F_{2t} \,\mathcal U_{t \epsilon}\, B(0) \right].
\end{eqnarray}
Averaging over  elements of $\mathcal U_{t\epsilon}$ and utilizing the unentangled form of $B(0) = \mathbb I \otimes \mathcal O_2$ results in \cite{supplemental},
%\begin{eqnarray}
%\nonumber & \overline{C_4(t)} = \Tr \left[  \left({\mathcal F_{1t}}^\dagger \otimes {\mathcal F_{2t}}^\dagger\right) A(t-1) ({\mathcal F_{1t}} \otimes {\mathcal F_{2t}})  B(0) \right.\\
%&\left. \times  \left({\mathcal F_{1t}}^\dagger \otimes {\mathcal F_{2t}}^\dagger\right) A(t-1) ({\mathcal F_{1t}} \otimes {\mathcal F_{2t}})  B(0) \right] \overline{{\mathcal U_{t\epsilon}}}^4 
%\end{eqnarray}
%As $B(0)$ is unentangled, being $\mathbb I \otimes O_2$, this simplifies to
\begin{equation}
\label{eq-c4-avg}
\overline{C_4(t)} = \sinc^4(\pi \epsilon)~\Tr \left[ A(t-1) B_{\text{loc}}(1) A(t-1) B_{\text{loc}}(1)\right],
\end{equation}
%where $  \overline{{\mathcal U_{t\epsilon}}}=\sinc(\pi \epsilon)$ and 
where $B_{\text{loc}}(1) = \left[ \mathbb I \otimes \mathcal F_{2t} \mathcal O_2 \mathcal F^{\dagger}_{2t} \right]$ denotes a local unitary evolution of the operator $B(0)$ {\it backward} in time. Importantly, it is also unentangled and hence the resultant correlator is again of the form in Eq.~(\ref{eq-otoc-2}). Recursive use of these approximations yields 
\begin{align}
\label{eq-c4-expression}
C_4(t) \approx \overline{C_4(t)}=\sinc^{4t}(\pi \epsilon) ~ \Tr (\mathcal O_1^2) ~\Tr (\mathcal O_2^2) .
\end{align}
Interestingly we did not have to average over the local operators $\mathcal F$, but we have verified that this also leads to the same result. The derivation also assumes that $\sinc(\pi \epsilon)$ is not very small, and hence is valid for $\epsilon \ll 1$, which is the case of weak interactions. In practice we find that the formula is good till $\epsilon \approx 0.2$ \cite{supplemental}.

If the operators $\Ob_j$ are diagonal in the same basis as the interaction, a situation in the standard map numerics we use, then 
the recursion is stopped after $(t-1)$ steps and the $C_4(t)$ is same as in Eq.~(\ref{eq-c4-expression}) with $t$ replaced by $(t-1)$.
The two point correlator $C_2(t)$ is approximated by the average over $\mathcal U_{t\epsilon}$ as well as the local unitary dynamics to result in
\begin{eqnarray}
%\begin{split}
\nonumber
C_2(t) &=&\Tr[ A^2(t) B^2(0)] \approx \frac{1}{N^2} \Tr \left(A(t-1)^2\right) \Tr \left(B(0)^2\right)  \\ 
& = &\Tr (\mathcal O_1^2) \Tr (\mathcal O_2^2).
%\end{split}
\end{eqnarray}
Thus within these approximations, the two point correlator $C_2(t)$ is trivially constant for all time and equal to $\Tr (\mathcal O_1^2) \Tr (\mathcal O_2^2)$, details are in  \cite{supplemental}.
Collecting the terms, the OTOC (for diagonal operators) is
\begin{equation}
\label{eq-otoc-formula}
C(t) = \Tr(\mathcal O_1^2)~ \Tr(\mathcal O_2^2) \left[ 1- \sinc^{4(t-1)}(\pi \epsilon)\right], \; t\geq 1.
\end{equation}
The OTOC of the random matrix model approaches saturation with an exponential decay with a rate $\mu_{RMT}(\epsilon)=-4 \ln |\sinc(\pi \epsilon)| \approx 2 \pi^2 \epsilon^2/3$ that is universal in the sense that it is independent of the choice of operators and depends only on the interaction.

For dynamical systems, with interaction propagator $U_b$,  $|\overline{U_{b}}|$ replaces $|\overline{\mathcal U_{t \epsilon}}|$, the average is found by considering the $q_j$ as random variables. Thus for quantum maps with the time between kicks being $\tau$ ($=1$ in the numerical results)
\beq
\label{eq-mu}
\mu(b)=-4 \ln \left | \int_{0}^1 d\xi_1 d \xi_2 e^{\frac{-i b V_{12}(\xi_1, \xi_2) \tau}{\hbar}} \right|\approx \frac{2 b^2 \tau^2}{\hbar^2}\overline{(\Delta V_{12})^2},
\eeq
which for the case of the coupled rotors considered here, with $V_{12}(\xi_1,\xi_2)=\cos[2 \pi(\xi_1 +\xi_2)]/(4 \pi^2)$, leads to Eq.~(\ref{eq:standmaprate}) and its validity is illustrated in Fig.~(\ref{fig-otoc-fft}). In the final approximation $\overline{(\Delta V_{12})^2}$
is the variance of the interaction.

\begin{figure}[!h]
\includegraphics[width=\linewidth]{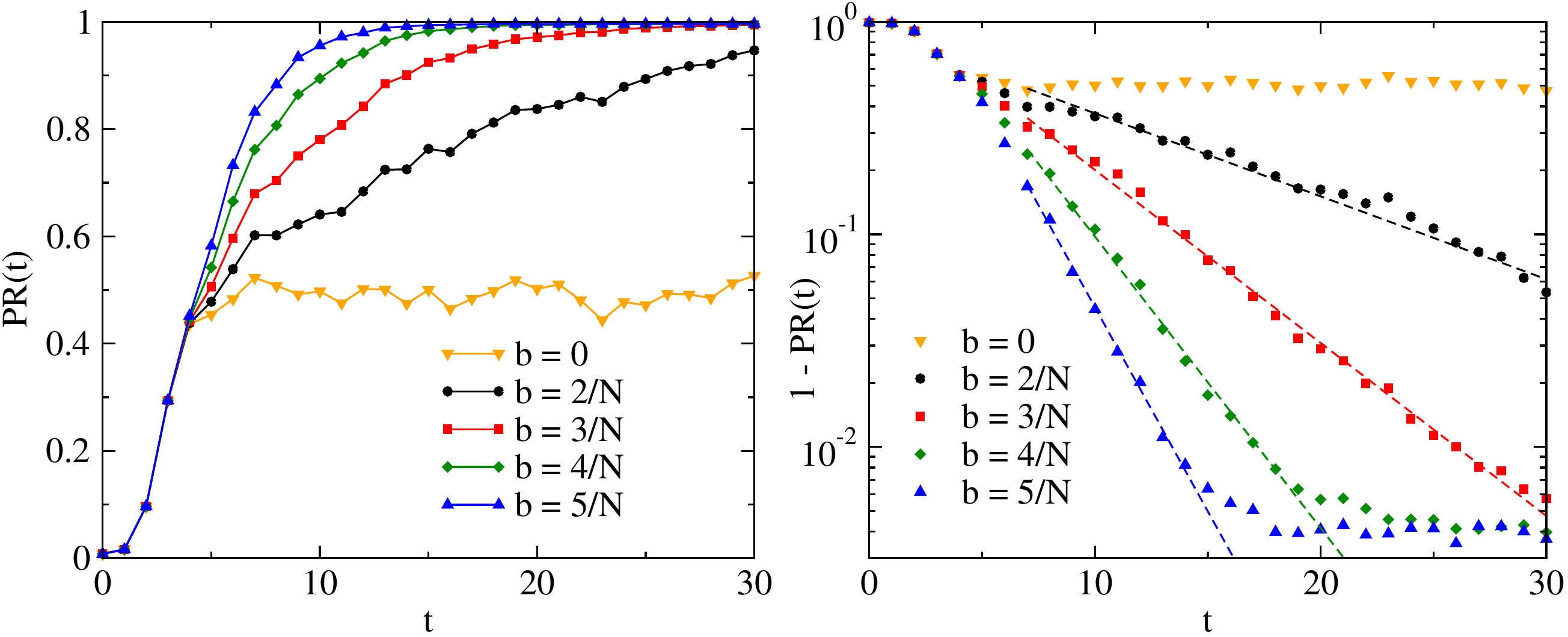}
\caption{(color online) (a) Time dependence of the phase-space participation ratio $\text{PR}(t)$ for the subsystem with $K_2 = 10$, evaluated for an initially localized product coherent state. Shown are several cases of the interaction $b$, all with $N = 256$ and $K_1=9$. (b) Shows the exponential relaxation of $\text{PR}(t)$ after the Ehrenfest time. \label{fig-pr-fft}}
\end{figure}

The picture of intra-subsystem scrambling giving way to inter-subsystem scrambling after the Ehrenfest time, is supported by 
studying the delocalization of initially localized (coherent) states in phase space: $|\psi(0)\kt = |q_1(0),p_1(0);q_2(0),p_2(0)\kt.$ The Husimi function 
$Q(t)_{q_1,p_1} = \Tr[ \rho_1(t)| q_1,p_1\kt\br q_1,p_1|]$ of the reduced density matrix of the subsystem of the time evolved state $U^n |\psi(0)\kt$ 
is helpful in visualizing the state in a given subsystem (Fig.~(\ref{fig:schemat}) is for the coupled rotors with $N=256$, $K_1=9$, $K_2=10$, $b=5/N$, $t_1=2$, $t_2=4$, $t_3=10$) and a measure of its delocalization is the participation ratio (PR) defined as,
%\begin{equation}
$\text{PR}(t) = 1/\sum_{q_1,p_1} Q(t)^2_{q_1,p_1}$,
%\end{equation}
with the maximum value of $1$ being the most delocalized in the subsystem phase-space. The participation ratio plotted in Fig.~(\ref{fig-pr-fft}) illustrates strikingly that during the first phase the delocalization is
independent of the interaction and essentially is that of the uncoupled system. It reaches the random matrix value pertaining to a random pure state,
namely $1/2$ at the Ehrenfest time before embarking on a interaction dependent second phase at which it relaxes to almost the maximum $1$ indicating global delocalization \cite{Arul-2004}. While the $\text{PR}(t)$ relaxation is also exponential as the OTOC, the rate is different and we do not yet have a precise estimate for it.

A similar scenario as above is found in preliminary many-body studies with spin-chains, which forms a natural extension. 
The study of mixed phases spaces, or when one of the subsystems is chaotic and other regular are of interest. Strong interaction 
results in the loss of subsystem identities and it is interesting that the RMT model described above overestimates the rate of
relaxation. Connections to relaxation rates such as the Ruelle-Pollicott resonances is also of interest.

\begin{acknowledgments}
R.P. acknowledges the SERB NPDF scheme (File No. PDF/2016/002900) for financial support.
\end{acknowledgments}

\bibliography{references.bib}

\end{document}

\title{Supplemental material for ``Scrambling in strongly chaotic weakly coupled bipartite systems: Universality beyond the Ehrenfest time-scale"}
%\title{Out-of-time-order correlators in bipartite chaotic systems: universal and non-universal features}
\author{Ravi Prakash} \email{raviprakash.sps@gmail.com}
\author{Arul Lakshminarayan}\email{arul@iitm.ac.in}
\affiliation{Indian Institute of Technology Madras, Chennai -- 600036, India}

\maketitle

In the supplemental material, we (i) give a detailed derivation for the OTOC using the RMT model, (ii) present data on its limitations with respect to large interaction strengths (iii) discuss the classical analogue of OTOC~-~Poisson bracket and numerically compare the classical and quantum Lyapunov exponents for the coupled standard maps, (iv) show the dynamics of a coherent state in the quantum phase space through Husimi functions illustrating the differences between pre- and post-Ehrenfest regimes.

Explicit form of the operators used in the main text are as follows.
In the position basis with $0 \leq n_j \leq N-1$
\[
\br n^\prime_j \left | U_{K_j} \right | n_j \kt  = \frac{1}{\sqrt{N}}\exp\left[ -i \frac{ N K_j}{2\pi}\cos\left[\frac{2\pi}{N}(n_j+\alpha)\right] \right] 
 \exp\left[ i \frac{\pi}{N}(n_j-n_j^\prime)^2 \right],
\]
while the interaction $U_b$ is a diagonal matrix with entries given by
\[
\br n_1' n_2' \left | U_b \right | n_1 n_2 \kt = \exp \left[ -i \frac{N b}{2\pi}\cos\left[\frac{2\pi}{N}(n_1+n_2+2\alpha) \right] \right] \delta_{n_1, n_1^\prime} \delta_{n_2, n_2^\prime}.
\]
The quantum phase $\alpha$ controls parity which we break with the choice of $\alpha=0.35$. 
Unlike most earlier studies, we choose to preserve time-reversal symmetry.

\section{Out-of-time order correlator}
We derive the expression for out-of-time-order correlator for bipartite system. The OTOC is defined through Eqs.~(1, 2).
% $C(t) = -(1/2) \Tr[A(t),B(0)]^2 = C_2(t) - C_4(t)$ where $C_2(t) = \Tr[A^2(t)B(0)^2]$ and $C_4(t) = \Tr[A(t)B(0)A(t)B(0)]$. A generic Hamiltonian can be written as, $H = H_1\otimes \mathbb I + \mathbb I \otimes H_2 + H_\text{int}$. 
For strong subsystem chaos, the local Floquet operators are modeled as being taken from circular unitary ensemble (CUE). The evolution for time $t$ is given by Eq.~(6) with $\mathcal F_{1j}$ and $\mathcal F_{2j}$ are independent members of CUE. The interaction $\mathcal U_{j\epsilon}$ consist of a unitary diagonal matrix $\langle m_1 n_1 | \mathcal U_{j \epsilon} | m_2 n_2 \rangle = \exp(2\pi i \epsilon \xi_{m_1n_1})\delta_{m_1 m_2} \delta_{n_1 n_2}$ with $\xi_k$ being a uniform random in $[-1/2, 1/2]$ \cite{Arul-PRL-2016} and independent for each time $j$.
%$\mathcal U^t \approx \mathcal U^{(t)} = \prod \mathcal F_1 \otimes \mathcal F_2 \mathcal U_\epsilon$ where $\mathcal F_j = \exp(-iH_j t/\hbar)$ and $\mathcal U_\epsilon = \exp(-iH_\epsilon t/\hbar)$. In the following derivation, we have $\mathcal F_j$ from CUE and $\langle i_1 i_2 | U_\epsilon | j_1 j_2\rangle = \exp(2\pi i \epsilon \xi_{i_1 i_2}) \delta_{i_1 j_1}\delta_{i_2 j_2}$ with $-1/2 <\xi < 1/2$ being a random variable. 

For the two-point function, $C_2(t)$, for $t>1$, we get after averaging over $\mathcal U_{t\epsilon}$,
\begin{eqnarray}
\nonumber C_2(t) &=& \Tr[ A^2(t) B^2(0)] \\
\approx \overline{C_2(t)} &=& \Tr \left[ \left({\mathcal F_{1t}}^\dagger \otimes {\mathcal F_{2t}}^\dagger\right) A(t-1)^2 \left( \mathcal F_{1t} \otimes \mathcal F_{2t} \right) \overline{\mathcal U_{t\epsilon} B(0)^2 \mathcal U_{t \epsilon}^{\dagger}} \right],
\end{eqnarray}
where $A(t) = \mathcal U_{t \epsilon}^\dag (\mathcal F_{1t}^\dag \otimes \mathcal F_{2t}^\dag) A(t-1) (\mathcal F_{1t} \otimes \mathcal F_{2t}) \mathcal U_{t \epsilon}$ and 
\begin{equation}
\label{eq-B-avg-Uep}
\left \langle i_1 i_2 \left| \overline{\mathcal U_{t\epsilon} B(0)^2 \mathcal U_{t \epsilon}^{\dagger}}  \right|j_1 j_2 \right\rangle  = \left< i_1 i_2 \left | B(0)^2 \right |j_1 j_2 \right> \overline{\langle j_1 j_2|\mathcal U_{t\epsilon}^\dagger| j_1 j_2\rangle \langle i_1 i_2|\mathcal U_{t\epsilon}| i_1 i_2\rangle}.
\end{equation}
The last term can be simplified to,
\begin{equation}
\label{eq-avg-Uep}
\overline{\langle j_1 j_2|\mathcal U_{t\epsilon}^\dagger| j_1 j_2\rangle \langle i_1 i_2|\mathcal U_{t\epsilon}| i_1 i_2\rangle} = \sinc^2(\pi \epsilon) (1-\delta_{i_1j_1} \delta_{i_2j_2}) + \delta_{i_1j_1} \delta_{i_2j_2}.
\end{equation}
Further, we average over the local random matrices $\mathcal F_{1t}$ and $\mathcal F_{2t}$:
\begin{align}
\label{eq-c2-avg}
\overline{C_2(t)} = \sum_{i_1, i_2,\ldots, l_1 l_2} \Big[
\overline{
\langle j_1 | \mathcal F_1^\star | i_1 \rangle
\langle k_1 | \mathcal F_1 | l_1 \rangle
}~
\overline{
\langle j_2 | \mathcal F_2^\star | i_2 \rangle
\langle k_2 | \mathcal F_2 | l_2 \rangle
}
\langle j_1j_2 | A(t-1)^2 | k_1 k_2 \rangle
\langle l_1l_2 | \overline{\mathcal U_{t\epsilon} B(0)^2 \mathcal U_{t \epsilon}^{\dagger}} | i_1 i_2 \rangle
\Big].
\end{align}
 Substituting $\overline{\langle j | \mathcal F | k \rangle \langle l | \mathcal F^\star | m \rangle} = \delta_{jl}\delta_{km}/N$ \cite{Collins-2006} in Eq.~(\ref{eq-c2-avg}), results in,
\begin{eqnarray}
\label{eq-c2-expression}
\overline{C_2(t)} = \frac{1}{N^2} \Tr \left(A(t-1)^2\right) \Tr \left(B(0)^2\right)
&=& \Tr (\mathcal O_1^2) \Tr (\mathcal O_2^2).
\end{eqnarray}

Similarly for the four-point function, we write $C_4(t)$ as,
\begin{eqnarray}
\label{eq-c4-2}
 C_4(t) & =\Tr \left[ \mathcal U_{t \epsilon}^\dagger (\mathcal F_{1t}^\dagger \otimes \mathcal F_{2t}^\dagger) A(t-1) (\mathcal F_{1t} \otimes \mathcal F_{2t}) \mathcal U_{t \epsilon} B(0)  \mathcal U_{t \epsilon} ^\dagger (\mathcal F_{1t}^\dagger \otimes \mathcal F_{2t}^\dagger) A(t-1) (\mathcal F_{1t}\otimes \mathcal F_{2t}) \,\mathcal U_{t \epsilon}\, B(0) \right].
\end{eqnarray}
Averaging over elements of $\mathcal U_{t\epsilon}$, results in
\begin{eqnarray}
\label{eq-c4-3}
\nonumber \overline{C_4(t)} &=& \Tr \Big[  \left({\mathcal F_{1t}}^\dagger \otimes {\mathcal F_{2t}}^\dagger\right) A(t-1) ({\mathcal F_{1t}} \otimes {\mathcal F_{2t}})  B(0)   \left({\mathcal F_{1t}}^\dagger \otimes {\mathcal F_{2t}}^\dagger\right) A(t-1) ({\mathcal F_{1t}} \otimes {\mathcal F_{2t}})  B(0) \\
&& \times \overline{\exp(2\pi i \epsilon (\xi_{i_1 i_2} + \xi_{j_1j_2} - \xi_{k_1 k_2} -\xi_{l_1 l_2}))} \Big].
\end{eqnarray}
The last term gives $\sinc^4(\pi \epsilon)$ to the leading order in $N$. As $B(0)=\mathbb I \otimes \mathcal O_2$ is unentangled, this simplifies to
\begin{align}
\label{eq-c4-1}
\overline{C_4(t)} & = \sinc^4(\pi \epsilon) \Tr[A(t-1) B_{\text{loc}}(1) A(t-1) B_{\text{loc}}(1)],
\end{align}
where $B_\text{loc}(1) = (\mathcal F_1 \otimes \mathcal F_2) B(0) (\mathcal F_1^\dag \otimes \mathcal F_2^\dag)  = (\mathbb I \otimes \mathcal F_{2t} \mathcal O_2 \mathcal F_{2t}^\dag)$ represents the local evolution of the operator {\it backward} in time. 
Since $B_\text{loc}$ is also unentangled and hence the resultant correlator is again of the form in Eq.~(2). Recursive use of these approximations yields 
\begin{align}
\label{eq-c4-appendix}
C_4(t) \approx \overline{C_4(t)}=\sinc^{4t}(\pi \epsilon) ~ \Tr (\mathcal O_1^2) ~\Tr (\mathcal O_2^2) .
\end{align}
Interestingly, the leading order term is independent of local operators $\mathcal F$ so we did not have to average over them. The lower order terms include the $\mathcal F$ but do not contribute significantly even after averaging over local unitary operators.

Thus, substituting Eq.~(\ref{eq-c4-appendix}) and Eq.~(\ref{eq-c2-expression}) in Eq.~(1) of the main text, we get final expression for OTOC,
\begin{equation}
C(t) = \Tr(\mathcal O_1^2) ~ \Tr(\mathcal O_2^2) [1-\sinc^{4t}(\pi \epsilon)],
\end{equation}
and the relaxation rate $\mu$ can be written as,
\begin{equation}
\label{eq-muep}
\mu(\epsilon) = -4 \ln \left[\sinc(\pi \epsilon) \right].
\end{equation}

\section{Validity of relaxation rate}
Beyond Ehrenfest time, the OTOC approaches saturation values with exponentially decaying rate. The OTOC in this regime can be modeled by random matrix and therefore we call it RMT phase. The relaxation rate depends only on the interaction and is given by Eq.~(12). For coupled kicked rotor and the RMT model the rate is given by Eq.~(5) and Eq.~(\ref{eq-muep}) respectively.

The relaxation rate can be used to find relation between parameters $b$ and $\epsilon$. When relaxation rates are equal for both quantum kicked rotor and RMT model, we get from Eqs.~(5, \ref{eq-muep}),
\begin{eqnarray}
\nonumber
\ln \left | J_0 \left(\frac{Nb}{2\pi}\right) \right|^{-4} &=& \ln |\sinc(\pi \epsilon)|^{-4} \\
\text{or}~ \frac{N^2 b^2}{4\pi^2} &=& \frac{2\pi^2 \epsilon^2}{3}
\end{eqnarray}
for $b,\epsilon \ll 1$. This simplifies to,
\begin{align}
\label{eq-ep-b}
\epsilon = \sqrt{\frac{3}{8}} \frac{Nb}{\pi^2},
\end{align}

Same relation can be obtained by considering the operator $U_b$ or $\mathcal U_\epsilon$ as perturbation to the uncoupled system. The relation can be computed by equating the parameter $\Lambda = v^2/D^2$ for both cases, where $v^2$ is the off-diagonal variance of elements of perturbation matrix when measured in the basis of unperturbed system, and $D$ is the mean level spacing in the spectrum \cite{Arul-PRL-2016}.

\begin{figure}[!h]
\includegraphics[width=0.5\linewidth]{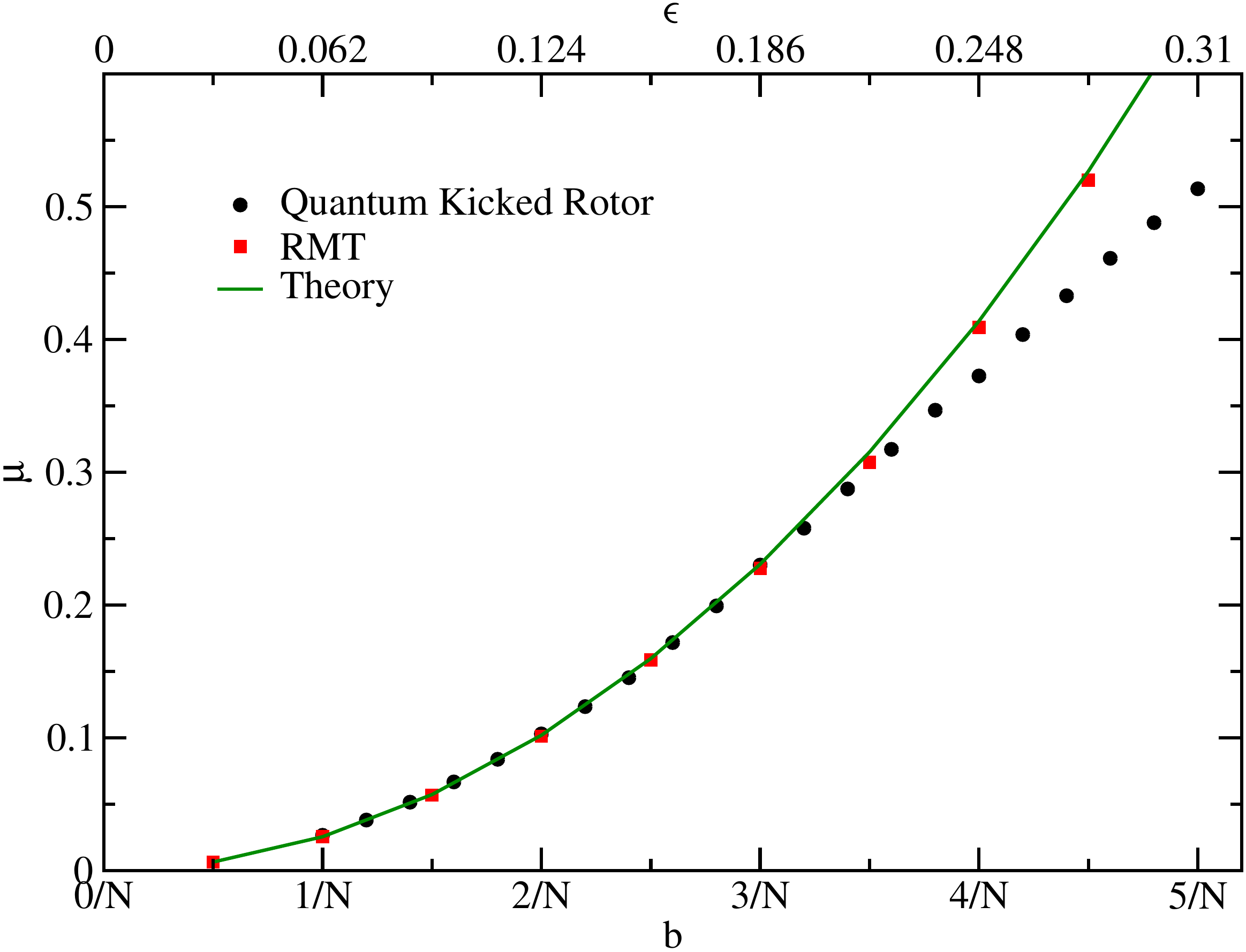}
\caption{Relaxation rate $\mu$ dependence on interaction in the RMT phase of OTOC. The plot with black circles shows the relaxation rate, $\mu(b)$ and with red squares represents the rate, $\mu(\epsilon)$ for the RMT model. The parameters $b$ and $\epsilon$ are scaled according to Eq.~(\ref{eq-ep-b}). The solid line correspond to Eq.~(12). We consider $N = 64$ for these plots. \label{fig-post-log-exponent}}
\end{figure}

We show in Fig.~({\ref{fig-post-log-exponent}}) how well the rates $\mu_{RMT}(\epsilon)$ and $\mu(b)$ agree with numerical simulations for a wide choice of couplings  $b$ and $\epsilon$. The $\epsilon$ is calculated from $b$ by Eq.~(\ref{eq-ep-b}). There is very good agreement with the theoretical values for even moderately large coupling, but we see systematic departures for very strong interactions when $b >\pi/N$. We speculate that at this level of interaction the subsystems do not randomize sufficiently to be modeled by random matrices at early times. It must be noted that no such deviations have been or can be seen in the properties of stationary states.

\section{OTOC for operators in same subspace}
When operators are in same subspace, {\it i.e.,} $A = \mathcal O_1 \otimes \mathbb I$ and $B = \mathcal O_2 \otimes \mathbb I$, the OTOC grow same as subsystem OTOC $(= \Tr[\mathcal O_1(t), \mathcal O_2(0)]^2)$, differ only by a multiplicative constant. For operators in same subspace, Eq.~(\ref{eq-c4-1}) can be written as,
\begin{align}
\nonumber
\overline{C_4(t)} & = N \sinc^{4t}(\pi \epsilon) \Tr[ A(0) B_\text{loc}(t) A(0) B_\text{loc}(t)] \\
& = N \sinc^{4t}(\pi \epsilon) \Tr[\mathcal O_1(0) \mathcal O_2(-t) \mathcal O_1(0) \mathcal O_2(-t)].
\end{align}
Here $B_\text{loc}(t) = (\mathcal F_{1}\mathcal O_2(0)\mathcal F_{1}^\dag \otimes \mathbb I)$ with $\mathcal F_1 = \mathcal F_{11} \mathcal F_{12} \ldots \mathcal F_{1t}$.
The last term is equal to the subsystem four-point function. The two-point function remain same and is $N$ times of subsystem two-point function. Thus OTOC for this case is approximately $N$ times of sub-system OTOC. Since four-point and two-point functions for subsystem are $O(1)$ and $O(N)$ implying $C(t) \approx C_2(t)$ for large $N$, The OTOC becomes constant after Ehrenfest time and we do not see a interaction dependent growth. This is also confirmed in Fig.~(2a).

\section{Classical Behavior - Poisson Brackets} \label{sec-poisson-bracket}
The OTOC exhibits an exponential growth (with rate $2\lambda_\text{L}$) in the Lyapunov phase $t < t_{EF}$ as shown in the inset of Fig.~(2a). The OTOC being a function of a commutator, has its classical analogue in terms of Poisson brackets. We show that the Poisson bracket grows exponentially in time for chaotic system with rate same as $2\lambda_\text{L}$. The corresponding classical function is Poisson bracket squared of equivalent classical functions $A = \cos(2\pi q_1))$ and $B = \cos(2\pi q_2)$,
\begin{align}
\label{eq-pb}
\nonumber
C_\text{cl}(t) & \propto  \left\{A(t), B(0)\right\}^2 \\
%\nonumber & \propto \left\{ \cos(2\pi q_1(t)), \cos(2\pi q_2(0))\right\}^2 \\
& = \sin^2(2 \pi q_1(t)) ~  \sin^2(2 \pi q_2(0))  ~ \left(\frac{\partial q_1(t)}{\partial p_2(0)}\right)^2.
\end{align}
For chaotic systems, the last term grows exponentially, $\partial q_1(t)/\partial p_2(0) \propto \exp(\lambda_\text{cl}t)$. Thus a qualitative expression for $C_\text{cl}$ can be written as,
\begin{align}
C_\text{cl}(t) = f(t) e^{2\lambda_\text{cl}t}.
\end{align}
To numerically estimate the Lyapunov exponent, we consider the ensemble average of $\lambda_\text{cl}$,
\begin{align}
\label{eq-lyapunov}
2\lambda_\text{cl}t  & = \overline{\log(C_\text{cl}(t))} - \overline{\log(f(t))}.
\end{align} 
Here the term $f(t) \propto \sin^2(2\pi q_1(t))$ fluctuates with time. Therefore $\overline{\log(f(t))}$ remains constant. Thus $\overline{\log(C_\text{cl}(t))}$ increases linearly in time with slope being equal to twice of the Lyapunov constant.

\begin{figure}[!h]
\includegraphics[width=0.6\linewidth]{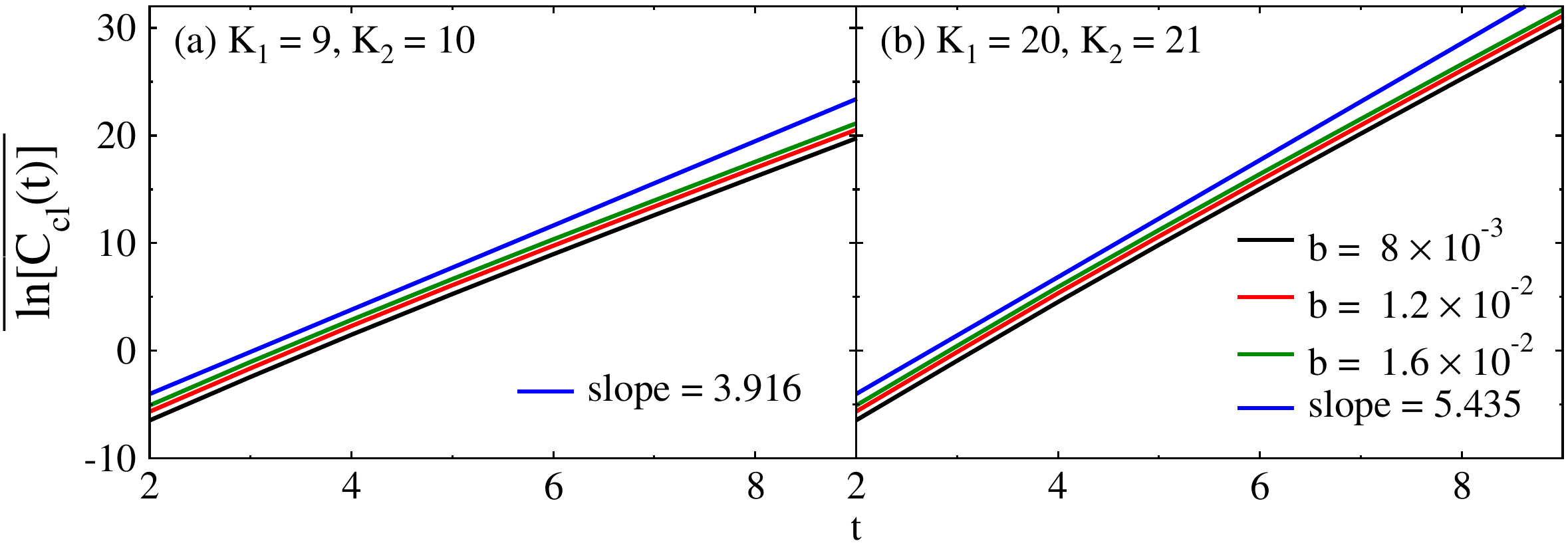}
\caption{The segment from $2\leq t\leq 5$ is considered for linear fitting. All curves represented by several values of $b$ exhibits similar slope. The slope is  $3.916$ for $K_1 = 9, K_2 = 10$ and $5.435$ for $K_1 = 20, K_2 = 21$. Here we have taken average over $1,00,000$ randomly distributed phase space points.\label{fig-poisson-bracket}}
\end{figure}

For numerics, let  $M_n$ represents the Jacobian at $n$-th kick {\it i.e.,} $\langle j | M_n | k \rangle = \partial x_j(n-1)/\partial x_k(n-1)$, where $(x_1, x_2, x_3, x_4) \equiv (p_1, q_1, p_2, q_2)$. The above mentioned term $\partial q_1(t)/\partial p_2(0)$ in Eq.~(\ref{eq-pb}) can be numerically evaluated by computing the element $\langle 2|J|3\rangle$ of matrix  $J$,
\begin{align}
J = M_{t-1} \ldots M_1 M_0.
\end{align}
%where $M_n$ the Jacobian mapping the perturbations in $4$-dimensional space at time $(n-1)$ to time $n$ and is given by,
%\begin{align}
%\left < j \left | M_n \right | k \right >= \frac{\partial\xi_j(n)}{\partial\xi_k(n-1)}.
%\end{align}
%Here $\xi_1, \xi_2, \xi_3$ and $\xi_4$ symbolically represents $p_1, q_1, p_2$ and $q_2$ respectively.
We numerically evaluate the right hand side of Eq.~(\ref{eq-lyapunov}) after each kick.
In numerical simulations, The initial values of $p_1(0), q_1(0), p_2(0), q_2(0)$ are taken from uniform random distribution in range $(0,1)$. We consider an ensemble of $1,00,000$ such initials. The Lyapunov exponent is evaluated in Fig.~(\ref{fig-poisson-bracket}).

The classical Lyapunov exponent is compared with the quantum exponent obtained from OTOC in the Table \ref{table-lambda-s}. The quantum exponents are in good agreements with classical ones.
\begin{table}[h]
\begin{ruledtabular}
\begin{tabular}  {l l l l}
%\toprule
& $\bm{2\lambda_\text{\bf L} (N = 64)}$ &$\bm{2\lambda_\text{\bf L} (N = 256)}$ & $\bm{2\lambda_\text{\bf cl}}$\\
\\
$\bm{K_1 = 9,}$&  $3.91\pm0.01$ & $4.00\pm 0.02$ & $3.916$\\
$\bm{K_2 = 10}$&$(t_E=2.584)$& $(t_E=3.445)$&  \\
\\
%$\bm{K_1 = 19,}$& $4.98\pm0.05$ & $4.57\pm 0.09$ &$5.341$\\
%$\bm{K_2 = 20}$ &$(t_E=1.806)$ & $(t_E=2.408)$ & \\
%\\
$\bm{K_1 = 20,}$& $5.030\pm0.06$ & $5.41\pm O(10^{-4})$ & $5.435$\\
$\bm{K_2 = 21}$ &$(t_E=1.769)$ & $(t_E=2.356)$ & 
\end{tabular}
\end{ruledtabular}
\caption{The comparison of rate $\lambda_\text{L}$ and $\lambda_\text{cl}$ for various combinations of $N$ and $k$. Note that the slope of linear part in inset of Fig.~(2) and in Fig.~(\ref{fig-poisson-bracket}) correspond to  $2\lambda_\text{L}$ and $2\lambda_\text{cl}$ respectively.  \label{table-lambda-s}}
\end{table}

\section{Scrambling in phase space}
In support of the fact that the dynamics is intra-subsystem until Ehrenfest time followed by inter-subsystem mixing, we numerically study the reduced Husimi function.
%, $Q(t)_{q,p} = $, for a coherent state, $|q_0, p_0; q_0,p_0\rangle$ localized at $(q_0, p_0) = (0.7, 0.3)$ in both subspaces. The lowest energy eigenstate of Harper Hamiltonian $H =  2 - (T_p + T_p^\dag + T_q + T_q^\dag)/2$ can be taken as the coherent state $|0,0\rangle$ \cite{Saraceno-1990}. 
The Husimi function for the coupled system $H(t)_{q_1,p_1; q_2,p_2}$ and state $|\psi(t) \rangle$ is given by,
\begin{align}
H(t)_{q_1,p_1; q_2,p_2} = \langle q_1,p_1;q_2,p_2 | \rho(t) | q_1,p_1;q_2,p_2 \rangle,
\end{align}
where $\rho(t) = |\psi(t) \rangle \langle \psi(t)|)$ is the density matrix for the normalized wave function and $| q_1,p_1;q_2,p_2 \rangle = |q_1,p_1\rangle \otimes |q_2,p_2\rangle$ corresponds to coherent state localized at $(q_1,p_1;q_2,p_2)$. %For the plots shown in Fig.~(\ref{fig-husimi}), we have $\left | \psi(0) \right > = | q_0,p_0;q_0,p_0 \rangle$ with $(q_0,p_0) = (0.7, 0.3)$.

Since the phase space for the coupled system is four dimensional, we can't visualize the dynamics. We instead observe the dynamics of one of the sub-system using, what is known as \textit{reduced Husimi function} \cite{Arul-2004}. For sub-system, $B$ the reduced Husimi function is given by,
\begin{align}
Q(t)_{q,p} = \langle q,p | \rho_B(t) | q,p \rangle,
\end{align}
where $\rho_B = \Tr_A \rho$ is the reduced density matrix, obtained by tracing over first subsystem,
%\begin{align}
%\left<j_2 \left | \rho_B \right| k_2 \right> = \sum_{j_1} \left < j_1, j_2 \left | \rho \right | j_1, k_2 \right > 
%\end{align}

\begin{figure} [!h]
\includegraphics[width=0.5\linewidth]{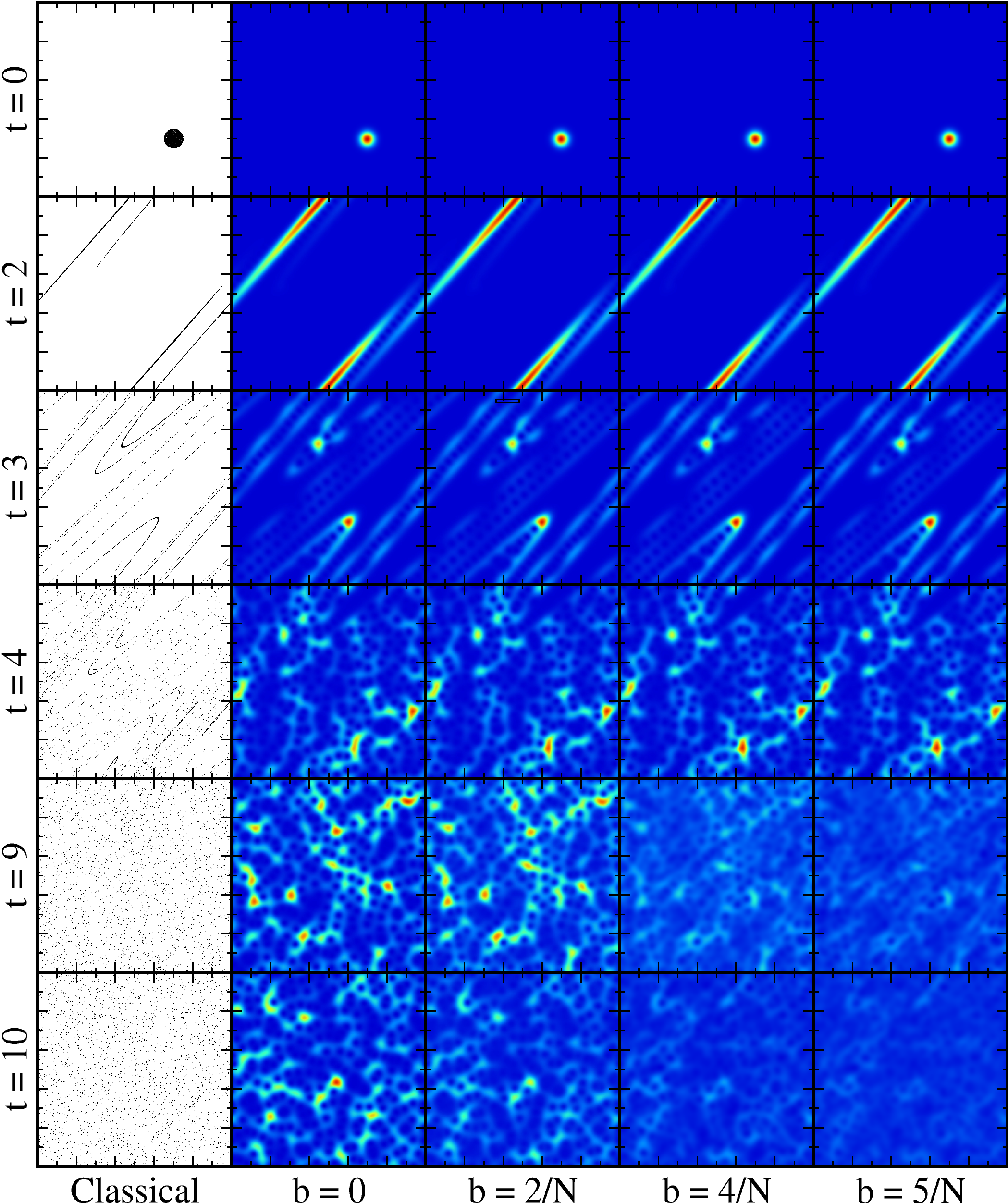}
\caption{Husimi phase-space representation of the subsystem state, $Q_{q,p}(t)$, with $K_2 = 10$, evolving under the dynamics of the coupled kicked rotor ($K_1=9$, $N=256$). The initial state is localized at $(q_0, p_0) = (0.7, 0.3)$ in both subspaces. The state looks identical at a fixed instant of time for several values of $b$ till the Ehrenfest time $t_E \approx 3.5$, indicating intra-subsystem dynamics, but starts getting smoothened out after the Ehrenfest time, faster for larger $b$, suggesting existence of inter-subsystem scrambling after the Ehrenfest time.\label{fig-husimi}}
\end{figure}

The ground state of the Harper Hamiltonian, $H = 2 - (T_q + T_q^\dagger + T_p + T_p^\dagger)/2$ can be taken as the coherent state $|0,0 \rangle$ for the subsystem \cite{Saraceno-1990}. We shift the ground state using position and momentum translation to get any state $\left| n,m \right > = T_p^m T_q^n \left |0,0 \right >$ localized at $(m,n)$

We show in Fig.~(\ref{fig-husimi}) the phase plot using reduced Husimi function for a coherent state, $|\psi(0) \rangle = |q_0, p_0; q_0,p_0\rangle$ localized at $(q_0, p_0) = (0.7, 0.3)$ in both subspaces. We observe that the Husimi function looks independent of $b$ till Ehrenfest time and start becoming flatter as $b$ increases after Ehrenfest time.

\bibliography{references.bib}